\def \SAIT #1 #2 {{\em Mem.\ Soc.\ Astron.\ It.\/} {\bf #1}, #2}
\def \MESS #1 #2 {{\em The Messenger\/} {\bf #1}, #2}
\def \ASTRNACH #1 #2 {{\em Astron. Nach.\/} {\bf #1}, #2}
\def \AAP #1 #2 {{\em Astron. Astrophys.\/} {\bf #1}, #2}
\def \AAL #1 #2 {{\em Astron. Astrophys. Lett.\/} {\bf #1}, L#2}
\def \AAR #1 #2 {{\em Astron. Astrophys. Rev.\/} {\bf #1}, #2}
\def \AAS #1 #2 {{\em Astron. Astrophys. Suppl. Ser.\/} {\bf #1}, #2}
\def \AJ #1 #2 {{\em Astron. J.\/} {\bf #1}, #2}
\def \ANNREV #1 #2 {{\em Ann. Rev. Astron. Astrophys.\/} {\bf #1}, #2}
\def \APJ #1 #2 {{\em Astrophys. J.\/} {\bf #1}, #2}
\def \APJL #1 #2 {{\em Astrophys. J. Lett.\/} {\bf #1}, L#2}
\def \APJS #1 #2 {{\em Astrophys. J. Suppl.\/} {\bf #1}, #2}
\def \APSS #1 #2 {{\em Astrophys. Space Sci.\/} {\bf #1}, #2}
\def \ASR #1 #2 {{\em Adv. Space Res.\/} {\bf #1}, #2}
\def \BAIC #1 #2 {{\em Bull. Astron. Inst. Czechosl.\/} {\bf #1}, #2}
\def \JSQRT #1 #2 {{\em J. Quant. Spectrosc. Radiat. Transfer\/} {\bf #1}, #2}
\def \MN #1 #2 {{\em Mon. Not. R. Astr. Soc.\/} {\bf #1}, #2}
\def \MEM #1 #2 {{\em Mem. R. Astr. Soc.\/} {\bf #1}, #2}
\def \PLR #1 #2 {{\em Phys. Lett. Rev.\/} {\bf #1}, #2}
\def \PASJ #1 #2 {{\em Publ. Astron. Soc. Japan\/} {\bf #1}, #2}
\def \PASP #1 #2 {{\em Publ. Astr. Soc. Pacific\/} {\bf #1}, #2}
\def \NAT #1 #2 {{\em Nature\/} {\bf #1}, #2}
\def\bge{\begin{equation}}
\def\ede{\end{equation}}

\documentstyle[twoside]{memsait}
\input epsf.sty
\begin{opening}
\title{CAN GALACTIC NUCLEI ACTIVITY BE EXPLAINED BY GLOBULAR CLUSTER MASS ACCRETION? }
\author{ROBERTO CAPUZZO--DOLCETTA$^1$}
\institute{$^1$Institute of Astronomy, University La Sapienza,
Roma, Italy}
\date{} 
\end{opening}

\begin{document}

\oddpagefooter{}{}{} 
\evenpagefooter{}{}{} 
\bigskip

\begin{abstract}
 Globular clusters can decay to the center of their mother galaxy
carrying a quantity of mass sufficient  to sustain the
gravitational activity of a small pre--existing nucleus and to accrete it
in a significant way. The AGN luminosity and lifetime depend on the
characteristics of the globular cluster system and it is quite 
insensitive to the nucleus' initial mass.
\end{abstract}

\section{Introduction}
It is commonly accepted  that the AGN emission is due  to gravitational
release of energy by mass falling onto a super--compact object.
Most of the attention is presently devoted to the geometry of the problem 
and to physical details in the attempt to explain some spectral 
characteristics and peculiarity of particular objects, rather than
to answer to the fundamental question: \\
\centerline{{\it where does the accreting mass come from?}}
\par In this note, which is part of a more extensive work described 
elsewhere (Capuzzo--Dolcetta, 1993; Capuzzo--Dolcetta, 1996) I try to give 
an answer to this question, carrying evidence 
that spherical mass accretion on a compact object in form of dynamically
decayed globular clusters in a (triaxial) elliptical galaxy can be the source of
the power released by AGNs without invoking ad hoc assumptions.
\section{Globular cluster spatial distribution in galaxies}
It seems quite ascertained that
the radial distribution of globular clusters is lees peaked than that of
bulge stars.  Recent WFPC2
HST observations of 14 elliptical galaxies 
(Forbes et al. 1996) confirm the general trend of 
flattening of the cluster distribution within $~ 2.5$ Kpc from the centre. 
This difference between the  bulge and the globular cluster distributions
may be  ab initio or  due  to 
 evolution of the cluster system  distribution.  This latter hypothesis
is  clearly the most ``economical''  and it has been positively checked
by Capuzzo--Dolcetta (1993).  Here I would point the attention to one of
the consequences that the evolution of the globular cluster system
(GCS) has in terms of mass carried to the galactic centre.

\section{The globular cluster system evolution}
The details of the general physical model can
be found in Capuzzo--Dolcetta  (1996) (see also Pesce , Capuzzo--Dolcetta and Vietri 1992,  and
Capuzzo--Dolcetta 1993). 
\par\noindent Here I just recall that the GCS evolution
in a diskless galaxy is due to the actions of {\it dynamical friction}
by field stars and of the {\it tidal} interaction with a compact object 
 in the centre of the galaxy. 
The relevant time scales ($\tau_{df}$, $\tau_{tid}$) depend on
the individual cluster initial energy, mass and mass density as well as
on the galactic potential and velocity distribution plus 
 (of course) on the nucleus mass ($M_n$). The ratio of the two time scales

\begin{equation}
{\tau_{tid}\over \tau_{df}}=f(E)\sqrt{\overline{\rho}_h}{M\over M_n}
\end{equation}

shows a competition between the two processes: dynamical friction is
 effective until a nucleus is accreted in mass enough
to destroy the incoming  clusters ($\overline{\rho}_h$ is 
the cluster density averaged over its half--mass radius). This
suggests the possibility of a self--regulated nucleus formation in a galaxy.
Various points have to be investigated, among them:
\par (i) when  $\tau_{tid}$ and $\tau_{df}$ are sufficiently short to be
relevant in the life of a galaxy?
\par (ii) is it possible to build up a compact nucleus in the galactic centre
in the form of dynamically decayed globulars?
\par (iii) what fraction of the mass of frictionally decayed and 
tidally destroyed clusters is swallowed in the compact nucleus?
\par (iv) can a quasar--like emission be explained by such kind
of spherical accretion?
\par A definite quantitative answer to question (ii) will require an 
accurate modelization; the answers to (i), (iii) 
(as well as  to (iv)) have been given 
in Pesce et al.  (1992), Capuzzo--Dolcetta  (1993), Capuzzo--Dolcetta  (1996),
and will not be discussed  here. 
\par In what follows I limit myself  to give support to the positive answer to question (iv).

\section{The accretion model}
Qualitatively: a black hole of given mass $m_{bh}$ stays at the galactic centre
and swallows the surrounding stars which enter a destruction radius
$r_d=max (r_S,r_t)$ where $r_S$ is the Schwarzschild's radius and
$r_t$ the tidal--breakup radius

$$
r_t=\left({ {m_{bh}\over <M_*>} }\right)^{1\over 3} <R_*>,
$$

with $ <M_*> $  and  $<R_*>$  mass and radius of the typical star.
 The resulting rate of (spherical) mass accretion is $\dot m_{bh}$, 
which yields a gravitational luminosity 

$$
L=\dot m_{bh} \phi
$$

($\phi$ is the gravitational potential near $r_d$).
They 
crucially depend on the star density 
($\rho_*$) and velocity dispersion ($<v_*^2>^{1/2}$) around the black hole
through 

\bge
\dot m_* =-\sigma_*\rho_*<v_*^2>^{1/2} \\
\ede
\bge
\sigma_* = \pi r_d^2\left({  1+{ {Gm_{bh}\over r_d} 
\over { {1\over 2} <v_{*}^2> }} }\right) \\
\ede
\bge
\dot m_{bh} =-\dot m_* \\
\ede
\bge
L =\eta \dot m_{bh}c^2
\ede

where $\dot m_*$ is the rate of stellar mass swallowed in the black hole
and $\eta$ is an efficiency factor, of the order of $10\%$.
Of course the higher $\rho_*$ the higher $\dot m$
while a high $<v_*^2>^{1/2}$ favours nucleus accretion increasing the
capture time rate but, contemporarily, decreases the swallowing
cross section (3).
\par  Capuzzo--Dolcetta (1996) with the help of a  
model similar to the one now described
(but more detailed) showed that  the 
nucleus accretion rate can easily increase up to 
few $M_\odot yr^{-1}$ due to  stars  which belonged 
to frictionally decayed globular clusters and 
to tidally disrupted clusters.  This mass accretion rate is of the order 
of that needed to sustain
a quasar  activity
 (note that the high velocity
bulge stars  cannot provide  for more than 
$10^{-7}-10^{-6} M_\odot yr^{-1}$).

\section{Results}
Here we report of some results of a model where the GCS
is composed by clusters of same mass and the system has an initial
density distribution and velocity dispersion equal to those of  bulge stars'  in  a typical triaxial galaxy. 
\par
Figure 1 a) shows the time evolution of the nucleus luminosity 
for different choices of the initial black hole mass  $m_{bh0}$
($10$,  $10^2$,  $10^3$,  $10^6$ $M_\odot$). It is 
 remarkable how,
independently of  $m_{bh0}$ ,  the highest luminosity
reached, $L_{max}$, is similar ($L_{max}$ varies for a factor 
of $10$ when $m_{bh0}$ ranges between $10$ to $10^6$ $M_\odot$).
The nucleus luminosity has
a short super--Eddington burst, a factor $10^3$ brighter than the 
luminosity when a 
relatively slow  dimming phase starts (the noise in the figure is real, due to 
the graininess of the problem amplified by the non--linear coupling of 
source and sink term in the dynamical equations).  Of course  $L_{max}$  is 
attained later for smaller  values of $m_{bh0}$, for
it requires longer to accrete enough mass  onto the nucleus. 
In all the cases the nucleus starts brightening   at 
about $1$ Gyr.
At that age enough  massive globulars have frictionally decayed to
the centre and released stars to the black hole. A flatter increasing slope
follows , due to that less massive incoming clusters feed less the nucleus.
The nucleus continues  increasing its mass  and becomes, rather quickly,
an efficient tidal destroyer of clusters capable  (in some cases) to keep in its potential
well a large fraction of the dispersed cluster' stars, which are eventually
swallowed. 
Figure 1 b) shows that the central black hole mass stabilizes around a 
value ($\approx 5\times 10^8$ M$_\odot$ in this model) which is about the 
same irrespectively its initial value.
\begin{figure}
\epsfysize=8cm 
\hspace{2.5cm}\epsfbox{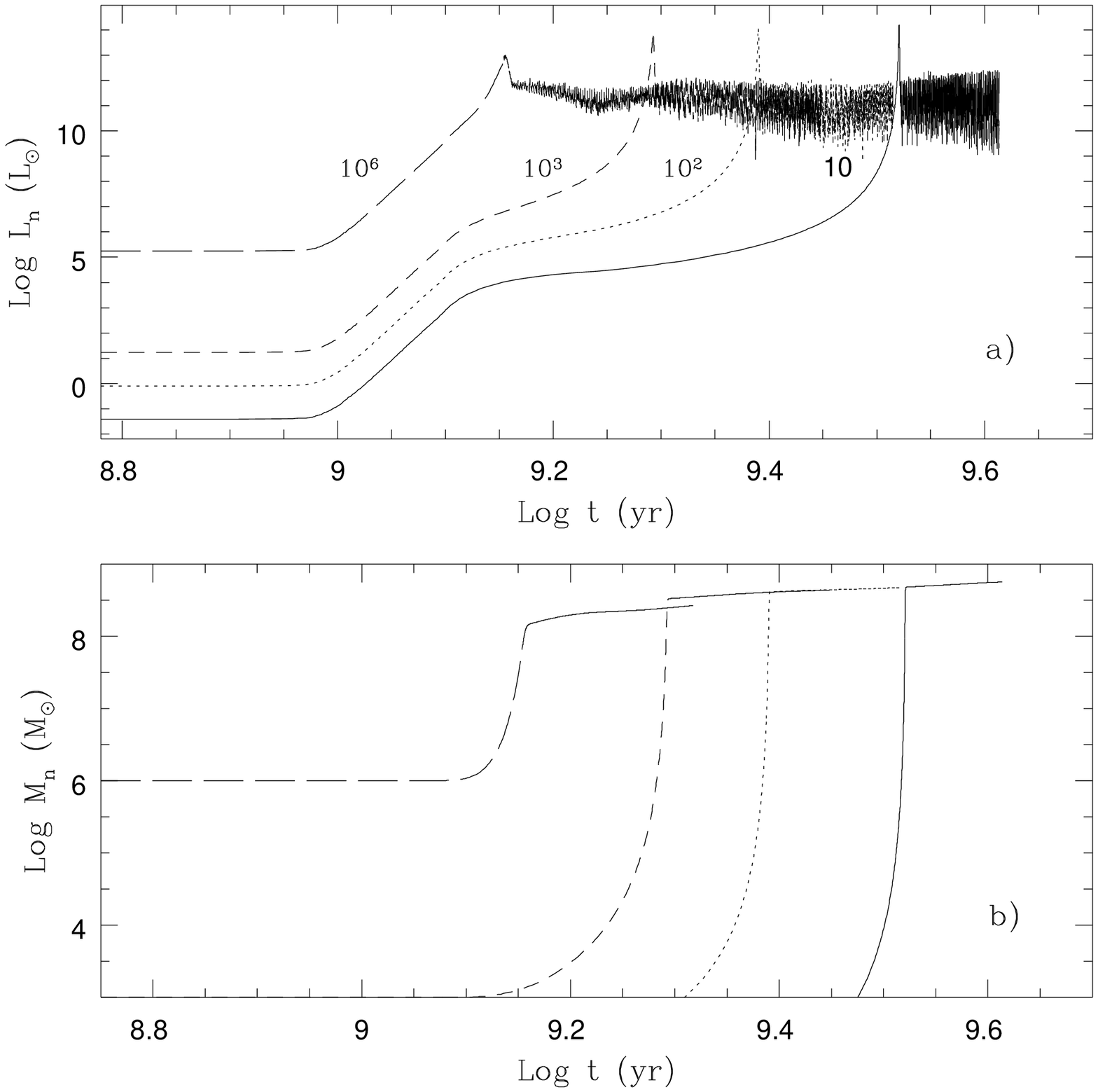} 
\caption{}{panel a): time evolution of  the  nucleus luminosity, for 
various initial nucleus masses (as labeled in $M_\odot$)
  ;  panel b):  time evolution of the nucleus mass, for the same initial
values  as in panel a).}
\end{figure}
 
\section{Conclusions}
I showed that a possible mechanism to accrete a compact nucleus 
placed in the centre
of a triaxial galaxy is the swallowing of surrounding stars, which belong
either to dynamically decayed clusters either to tidally destroyed ones. 
The quantitative relevance of this phenomenon, as well as its
time scales, depend of course on the orbital structure of the GCS
and on its mass and internal density spectrum: a "cold" system
composed by massive, dilute clusters is the most affected by both
dynamical friction and tidal erosion.
In the model here presented,  and
deeply discussed in Capuzzo--Dolcetta (1996), 
the peak of mass accretion rate and the final value of the nucleus mass
are remarkably 
independent (for a fixed typical globular cluster mass) of the initial
black hole mass. What changes is the age of occurrence of the 
resulting luminosity peak ($10^{13} \div 10^{14}$ L$_\odot$) 
and of the flattening of the mass growth curve. 
It is indeed found that the black hole mass grows rapidly
until it reaches a value of few $10^8$ M$_\odot$; after that the mass
accretion is much slower.


\end{document}